\documentstyle[emulateapj,epsf,apjfonts]{article}

\righthead{Astrometric and Light-travel Time Orbits of R Canis Majoris}
\lefthead{Ribas et al.}

\begin{document}

\title{Astrometric and Light-travel Time Orbits to Detect Low-mass Companions:
A Case Study of the Eclipsing System R Canis Majoris}

\author{Ignasi Ribas\altaffilmark{1,2}, Fr\'ed\'eric Arenou\altaffilmark{3},
and Edward F. Guinan\altaffilmark{1}} 

\altaffiltext{1}{Department of Astronomy \& Astrophysics, Villanova University,
Villanova, PA 19085, USA. E-mail: iribas@ast.villanova.edu,
edward.guinan@villanova.edu} 

\altaffiltext{2}{Departament d'Astronomia i Meteorologia, Universitat de
Barcelona, Av. Diagonal, 647, E-08028 Barcelona, Spain} 

\altaffiltext{3}{DASGAL, Observatoire de Paris, CNRS UMR 8633, 92195 
Meudon Cedex, France.  E-mail: Frederic.Arenou@obspm.fr}

\begin{abstract}
We discuss a method to determine orbital properties and masses of low-mass
bodies orbiting eclipsing binaries. The analysis combines long-term eclipse
timing modulations (light-travel time or LTT effect) with short-term,
high-accuracy astrometry. As an illustration of the method, the results of a
comprehensive study of Hipparcos astrometry and over a hundred years of eclipse
timings of the Algol-type eclipsing binary R Canis Majoris are presented. A
simultaneous solution of the astrometry and the LTTs yields an orbital period
of $P_{12}=92.8\pm1.3$~yr, an LTT semiamplitude of $2574\pm57$~s, an angular
semi-major axis of $a_{12}=117\pm5$ mas, and values of the orbital eccentricity
and inclination of $e_{12}=0.49\pm0.05$, and $i_{12}=91.7\pm4.7$~deg,
respectively. Adopting the total mass of R~CMa of
$M_{12}=1.24\pm0.05$~M$_{\odot}$, the mass of the third body is
$M_3=0.34\pm0.02$~M$_{\odot}$ and the semi-major axis of its orbit is
$a_3=18.7\pm1.7$~AU. From its mass, the third body is either a dM3-4 star or,
more unlikely, a white dwarf. With the upcoming microarcsecond-level
astrometric missions, the technique that we discuss can be successfully applied
to detect and characterize long-period planetary-size objects and brown dwarfs
around eclipsing binaries. Possibilities for extending the method to pulsating
variables or stars with transiting planets are briefly addressed.
\end{abstract}

\keywords{stars: individual (R~CMa) --- astrometry --- stars: fundamental
parameters --- binaries: eclipsing --- stars: late-type}

\section{Introduction}

Within the next decade several space astrometry missions, FAME and DIVA then
SIM and GAIA, capable of sub-milli-arcsecond to micro-arcsecond accuracy are
expected to be launched. One of the primary scientific goals of these missions
is the astrometric detection of low mass objects around nearby stars, including
brown dwarfs and Jupiter-sized planets. The detection of these objects will be
accomplished through the observation of the reflex motion of the host star
caused by the gravitational pull of the low-mass body. Although these missions
are capable of very high astrometric accuracies and can observe up to millions
of stars, their lifetimes are relatively short (2.5 to 5 yr). Thus, these space
missions are optimized to detect planets within the habitable zones of
late-type stars, but they could fail to detect (additional) planets with longer
periods. It is important, however, to secure a complete picture of the bodies
orbiting a star both from a pure census point of view and also to understand
the genesis and evolution of planetary systems. In addition, planets do not
necessarily remain within the habitable zone because of long-term chaotic
perturbations. As we know from our Solar System, the presence of massive
planets, such as Jupiter and Saturn, in distant orbits play a crucial role in
stabilizing the orbits of the inner planets. 

One effective way of extending the time baseline that permits the discovery of
long period exosolar planets or brown dwarfs is to use the light-travel time
(hereafter LTT) effect in eclipsing binaries. From this technique, the eclipses
act as an accurate clock for detecting subtle variations in the distance to the
object (this is analogous to the method used for discovering earth-sized
objects around pulsars, see Wolszczan \& Frail 1992). The periodic
quasi-sinusoidal variations of the eclipse arrival times have a very simple and
direct physical meaning: the total path that the light has to travel varies
periodically as the eclipsing pair moves around the barycenter of the triple
system. The amplitude of the variation is proportional both to the mass and to
the period of the third body, as well as to the sine of the orbital inclination. As
discussed by Demircan (2000), nearly 60 eclipsing binaries show evidence for
nearby, unseen tertiary components using LTT effects. A recent example of a
brown dwarf detected around the eclipsing binary V471~Tau using this method was
presented by Guinan \& Ribas (2001). Also this method is being employed in
selected low mass eclipsing binaries to search for extrasolar planets (Deeg et
al. 2000).

The primary advantages of using the LTT effect to detect third bodies in
eclipsing binaries are: {\em i)} The necessary photometry can be secured with
small telescopes using photoelectric or CCD detectors; {\em ii)} The number of
eclipsing binaries is large -- 4000 currently known in the Galaxy -- and this
number could increase very significantly when results from upcoming
astrometry and photometry (e.g., MONS, COROT, Eddington, Kepler) missions are
available; {\em iii)} For select eclipsing binaries (with sharp and deep
eclipses) the timings can be determined with accuracies as good as several
seconds; {\em iv)} The mass of the eclipsing pair can be known from
conventional spectroscopic and light curve analyses. A shortcoming of
the LTT method is that only upper limits to the mass and size of the orbit of
the tertiary component can be determined (the analysis yields the mass 
function\footnote{$f(M_3)=(M_3^3 \sin i_3^3)/(M_{12}+M_3)^2$}, $f(M_3)$, and 
$a_3 \sin i_3$). However, as it was demonstrated in the case of Algol (Bachmann 
\& Hershey 1975), the LTT analysis can be complemented with astrometry to yield
the orbital inclination and thus the actual mass and semi-major axis of the
third body. Furthermore, with the orbital properties ($P$, $e$ and $\omega$)
known from the LTT analysis, only a small fraction of the astrometric orbit
needs to be covered when using high-accuracy astrometry.

In this paper we present the results of the combined LTT analysis and Hipparcos
astrometry of the Algol-type eclipsing binary R~CMa. The residuals of over 150
eclipse timings extending from 1887 to 2001 show a periodic ($\sim$93 yr)
quasi-sinusoidal modulation. As previously shown by Radhakrishnan, Sarma, \&
Abhyankar (1984) and Demircan (2000), these variations are best explained by
the LTT effect arising from the gravitational influence of a third body. The
Hipparcos astrometry also shows the presence of small but significant
acceleration terms in the proper motion components explicable by the reflex
motion from a third body. Our study illustrates that with a well-defined LTT
effect, only a few years of accurate astrometry are needed to constrain the
orbital solution and determine the mass of the third body.

\section{Overview of R~CMa}\label{over}

R Canis Majoris (HD 57167, HR 2788, HIP 35487) is a bright ($V_{\rm max}=5.67$
mag), semi-detached eclipsing binary having an orbital period of 1.1359 days.
As pointed out by Varricatt \& Ashok (1999), R~CMa holds special status among
Algol systems in that it is the system with the lowest known total mass and
hosting the least massive secondary star. Since the discovery of its
variability in 1887 by Sawyer (1887), R~CMa has been frequently observed and
has well determined orbital and physical properties. The major breakthrough in
understanding the system came when Tomkin (1985) was able to measure the very
weak absorption lines of the faint secondary star and determine the masses of
the two stars from a double line radial velocity study. The analyses of its
light and radial velocity curves (see Varricatt \& Ashok 1999) show that this
system has a circular orbit and consists of a nearly spherical F0-1 V star 
($M_1=1.07\pm0.2$~M$_{\odot}$; $R_1=1.48\pm0.10$~R$_{\odot}$; $L_1/{\rm 
L}_{\odot}=5.78\pm0.38$) and a low mass, tidally-distorted K2-3 IV star 
($M_2=0.17\pm0.02$~M$_{\odot}$; $R_2=1.06\pm0.07$~R$_{\odot}$; $L_2/{\rm 
L}_{\odot}=0.43\pm0.10$). Moreover, nearly every photometric study indicates 
that the cooler star fills its inner Lagrangian surface. The relatively high 
space motions ($S=67$~km~s$^{-1}$) suggest that R~CMa is a member of the old 
disk population and thus a fairly old (5--7 Gyr) star (Guinan \& Ianna 1983).

The present state of the system is best explained as a low mass Algol system
that has undergone mass exchange and extensive mass loss. Asymmetries in its
light curves and subtle spectroscopic anomalies indicate that mass exchange and
loss are still continuing but at a much diminished rate compared to most Algol
systems. The very low mass of the secondary star and old disk age indicate that
R~CMa is near the end of its life as an Algol system. As in the case of all
Algol systems, the secondary star lies well above the main-sequence. However,
unlike most Algol systems, the primary star is too hot and overluminous for
observed mass. Moreover, a recent analysis of older photometry of R~CMa by
Mkrtichian \& Gamarova (2000) indicates that the F star is a low-amplitude
$\delta$~Scuti variable with a B light amplitude of 9 millimagnitudes and a
period of 68 minutes.

\section{Observations}

\subsection{Astrometry}\label{astro}

Hipparcos observed R~CMa between March 9, 1990 and March 5, 1993. There are 68
one-dimensional astrometric measurements corresponding to 35 different epochs
in the Hipparcos Intermediate Astrometric Data, which were obtained by the two
Hipparcos Data Reduction Consortia (33 measurements from FAST and 35 from
NDAC). The astrometric data can be obtained from CD-ROM 5 of the Hipparcos
catalog (ESA 1997). Unfortunately, the timespan of the Hipparcos observations
is much smaller than the orbital period of the tertiary component, and this
might eventually result in possible systematic errors in the orbital elements.
To further constrain the solution additional older ground-based positions must
be used. Indeed, Tycho-2 proper motions were computed by combining Tycho-2
positions and ground-based astrometric catalogs. For R~CMa, 17 epoch positions
of ground-based catalogs used for the Tycho-2 proper motion computation were
kindly made available to us by Dr. Urban and are listed in Table
\ref{tabastro}. These measurements span over one century and so the Tycho-2
proper motions can be understood as the combination of the true proper motion
and of a large fraction of the orbital motion. Consequently, the Tycho-2 proper
motion of R~CMa cannot itself be used in our analysis and only the individual
positions contain valuable orbital information. In contrast, a short-term
proper motion determination, such as the one computed around 1980 by Guinan \&
Ianna (1983), reveals itself to be very useful in constraining the astrometric
solution.

In the course of the astrometric data reduction of the Hipparcos data, a test
was applied to all the (apparently) single stars in order to check whether
their motion was significantly non-linear. Most likely, a significant curvature
of the photocenter motion is an indication of a possible duplicity.  As it
turns out, R~CMa is one of the 2622 ``acceleration'' solutions of the Double
and Multiple Stars Annex of the Hipparcos Catalogue, which provides a hint for
the presence of a third body, independently from the LTT effect.

\subsection{Photometry} \label{secphot}

R~CMa has a long baseline of eclipse timings that extend from the present back
to 1887. Most of the early eclipse times were determined from visual estimates.
Several period studies have been carried out. Early studies of times of minimum
light indicated a possible abrupt decrease in the period during 1914--15 (see
Dugan \& Wright 1939; Wood 1946; Koch 1960; Guinan 1977). However, as more
timings accumulated it became apparent that the long-term variations in the
(O--C)'s of the system are periodic and thus best explained by the LTT effect
produced by the presence of a third body. The analysis of Radhakrishnan et al.
(1984), with eclipse timings from 1887 to 1982, and Demircan (2000), who
includes timings up to 1998, make a strong case in support of the LTT scenario.

Our photoelectric eclipse timing observations extend the time baseline up to
early 2001. The observations were obtained with the Four College 0.8-m
Automatic Photoelectric Telescope located in Southern Arizona during 1995/96
and 2000/01. Differential photometry was carried out using $uvby$ Str{\"o}mgren
filter sets. The mid-times of primary minimum and the (O--C)'s for these are
given in Table \ref{tabtim}, along with the corresponding uncertainties. The
(O--C)'s were computed using a refined ephemeris determined from the analysis
in \S\ref{analys} (Eq. \ref{ephem}). Our observations were combined with those
compiled from the literature to yield a total of 158 eclipse timings from 1887
through 2001. The primary eclipse observations obtained from the literature are
also provided in Table \ref{tabtim}. Even though it is not explicitly mentioned
in any of the publications, the times listed are commonly assumed to be in the
UTC (coordinated universal time) scale. The timings were transformed from UTC
to TT (terrestrial time) following the procedure described in Guinan \& Ribas
(2001), which is based on the recommendations of Bastian (2000). The timings
listed in Table \ref{tabtim} are therefore HJD but in the TT scale.

The uncertainties of the individual timings are difficult to estimate. The
compilation of Radhakrishnan et al. (1984) (from which most of the timings in
Table \ref{tabtim} come) does not provide timing errors but only a relative
weighting factor related to the quality of the data and the observation
technique. We therefore adopted an iterative scheme to determine the actual
uncertainties by forcing the $\chi^2$ of the (O--C) curve fit (in \S
\ref{analys}) to be equal to unity. This rather arbitrary scale factor
determination is indeed justified because it ensures that the fitting algorithm
will yield realistic uncertainties for the orbital parameters of the system.
The individual timing errors are included in Table \ref{tabtim}. The
uncertainties yielded by the iterative scheme are about 10--13 minutes for
photographic timings and some 2--4 minutes for photoelectric timings, in both
cases reasonable figures given the characteristics of the two methods and the
shape of the eclipse.

The Hipparcos mission, in addition to observing accurate positions of R~CMa,
obtained a total of 123 photometric measurements, which are present in the
Hipparcos Epoch Photometric Data (ESA 1997). The phase coverage of the 
observations is not sufficient to determine an accurate primary eclipse timing
using conventional methods. As an alternative, we adopted the physical
information available to fit the entire light curve and derive a phase offset.
Also, since the observations span 3 yr and the system exhibits (O--C)
variations, the photometric dataset was split into two subgroups about January
1, 1991.  Then, using the photometric elements of Varricatt \& Ashok (1999), we
employed the Wilson-Devinney program (Wilson \& Devinney 1971) to run fits to
both light curves by leaving only a phase shift and a magnitude zero point as
free parameters. The fits were very satisfactory and we derived eclipse timings
for the mean epochs from the best-fitting phase offsets. The resulting two
timings, with uncertainties of around 100 s, are included in Table
\ref{tabtim}.

These two timings are significantly different as one would expect from the LTT
secular change of period during the 3-year duration of the Hipparcos
observations. More precisely, each date of observation could be approximated by
$T\approx T_{\circ}+P_{\circ} E+g E^2$, with $E$ being the variability cycle
number. So, the photometric ``acceleration'' term $g$ is a measure of the
departure from a linear ephemeris during the observation window (the Hipparcos
mission lifetime in this case). Since Hipparcos did not provide minimum
timings, we used an alternative method to fit the equation above. The epoch
measurements were folded using the reference epoch and period from ground-based
studies. Then, the quadratic term was estimated by minimizing the distance
between successive points of the folded light curve (string-length method).
The fit yielded a value of $g=2.09\cdot10^{-8}$ d, which leads to a cumulative
effect of $g E_{\rm max}^2=0.0193\pm0.0006$ d during the course of the
Hipparcos observations. This rough estimation gives a significant acceleration
term which is of the same order as that resulting from the long-term LTT
analysis (\S \ref{analys}) and indicates that the (O--C) variations attributed
here to the presence of a third body could, in principle, have been detected
through the Hipparcos photometric analysis alone.

\section{Analysis}\label{analys}

The expressions that describe the LTT effect as a function of the orbital
properties were first provided by Irwin (1952). In short, the time delay or
advance caused by the influence of a tertiary component can be expressed as:
\begin{equation} \label{eq1}
\Delta T = \frac{a_{12} \sin i_{12}}{c} \left[\frac{1-e_{12}^2}{1+e_{12} 
\cos \nu_{12}} \sin(\nu_{12} + \omega_{12}) + e_{12} \sin \omega_{12} \right]
\end{equation}
where $c$ is the speed of light, and $a_{12}$, $i_{12}$, $e_{12}$,
$\omega_{12}$, and $\nu_{12}$ are, respectively, the semi-major axis, the
inclination, the eccentricity, the argument of the periastron, and the true
anomaly (function of time) of the orbit of the eclipsing pair around the
barycenter. As is customary, the orbital inclination $i_{12}$ is measured
relative to the plane of the sky. The naming convention adopted throughout this 
paper uses subscripts 12 and 3 for the orbital parameters of the eclipsing 
pair and the tertiary component, respectively, around the common barycenter. 
Obviously, most of the parameters for the eclipsing pair's and the third 
body's orbits are identical (such as period, eccentricity, and inclination), 
but we use the subscript 12 because the actual measurements are made strictly 
for the brighter component of the long period system. Finally, subscript EB 
refers to the close orbit of the eclipsing binary.

The fit of Eq. (\ref{eq1}) to the timing data would provide a good estimation
of the a number of orbital and physical parameters of the system (see e.g.,
Guinan \& Ribas 2001). However, both the orbital semi-major axis and the mass
of the third body would be affected by a factor $\sin i$ that cannot be
determined from the LTT analysis alone. When the LTT analysis is combined with
astrometric data, all orbital parameters (including $i$ and even $\Omega$) can
be determined yielding a full description of the system.  The availability of
Hipparcos intermediate astrometry permits the fitting of the observations using
an astrometric model not accounted for in the standard Hipparcos astrometric
solution. In the particular case of R~CMa we have considered an orbital model
that has been convolved with the astrometric motion (parallax and proper
motion). The orbital motion produces the following effect on the coordinates:
\begin{eqnarray}
\Delta x &=& a_{12} \; \frac{1-e_{12}^2}{1+e_{12}\cos \nu_{12}}
[\cos(\nu_{12}+\omega_{12}) \sin\Omega_{12} + \nonumber \\ 
\label{astro1}
&&+\sin(\nu_{12}+\omega_{12}) \cos\Omega_{12} \cos i_{12}] \\
\Delta y &=& a_{12} \; \frac{1-e_{12}^2}{1+e_{12}\cos \nu_{12}}
[\cos(\nu_{12}+\omega_{12}) \cos\Omega_{12} - \nonumber \\
\label{astro2}
&&-\sin(\nu_{12}+\omega_{12})\sin\Omega_{12} \cos i_{12}]
\end{eqnarray}
In addition, since the Hipparcos measurements are unidimensional, the variation
of the measured abscissa $v$ on a great circle is:
\begin{eqnarray}
\Delta v & = & {{\partial v}\over{{\partial \alpha\cos\delta}}}
(\Delta\alpha\cos\delta+\Delta x) +
{{\partial v}\over{{\partial \delta}}} (\Delta\delta+\Delta x) +
{{\partial v}\over{{\partial \varpi}}} \Delta\varpi + \nonumber \\
\label{astro3}
&&{{\partial v}\over{{\partial \mu_\alpha\cos\delta}}}
\Delta\mu_\alpha\cos\delta +
{{\partial v}\over{{\partial \mu_\delta}}} \Delta\mu_\delta
\end{eqnarray}
where the astrometric components are $\alpha$, $\delta$ for the coordinates,
$\mu_\alpha$, $\mu_\delta$ for the proper motion and $\varpi$ for the parallax.

In addition to fitting the orbital and astrometric properties of the system, a
timing zero point and a correction to the orbital period of the eclipsing pair
(that could lead to a linear secular increase or decrease of the (O--C)'s) were
also considered. The initial values of the period and the reference epoch were
adopted from Varricatt \& Ashok (1999). 

The full set of observational equations includes those related to the timing
residuals (Eq. (\ref{eq1})) and those coming from the astrometric measurements
(Eqs. (\ref{astro1}), (\ref{astro2}), \& (\ref{astro3})). All these equations
were combined together and the 14 unknown parameters (5 for the astrometric
components -- $\alpha$, $\delta$, $\mu_\alpha$, $\mu_\delta$, $\varpi$ --, 7
for the orbital elements -- $a_{12}$, $e_{12}$, $\omega_{12}$, $i_{12}$,
$P_{12}$, $T^{\rm peri}_{12}$, $\Omega_{12}$ --, one for the reference epoch --
${T_{\circ}}_{\rm EB}$ --, and one for the period of the eclipsing system --
$P_{\rm EB}$ --) were recovered via a weighted least-squares fit as described
in, e.g., Arenou (2001) or Halbwachs et al. (2000). Note that the weights of
the individual observations were computed as the inverse of the observational
uncertainties squared and multiplied by the corresponding correlation factors.
The uncertainties adopted are given in Tables \ref{tabastro} and \ref{tabtim},
and in the Hipparcos Catalogue CD-ROM 5.

\begin{figure}
\figurenum{1}
\epsscale{0.50}
\plotone{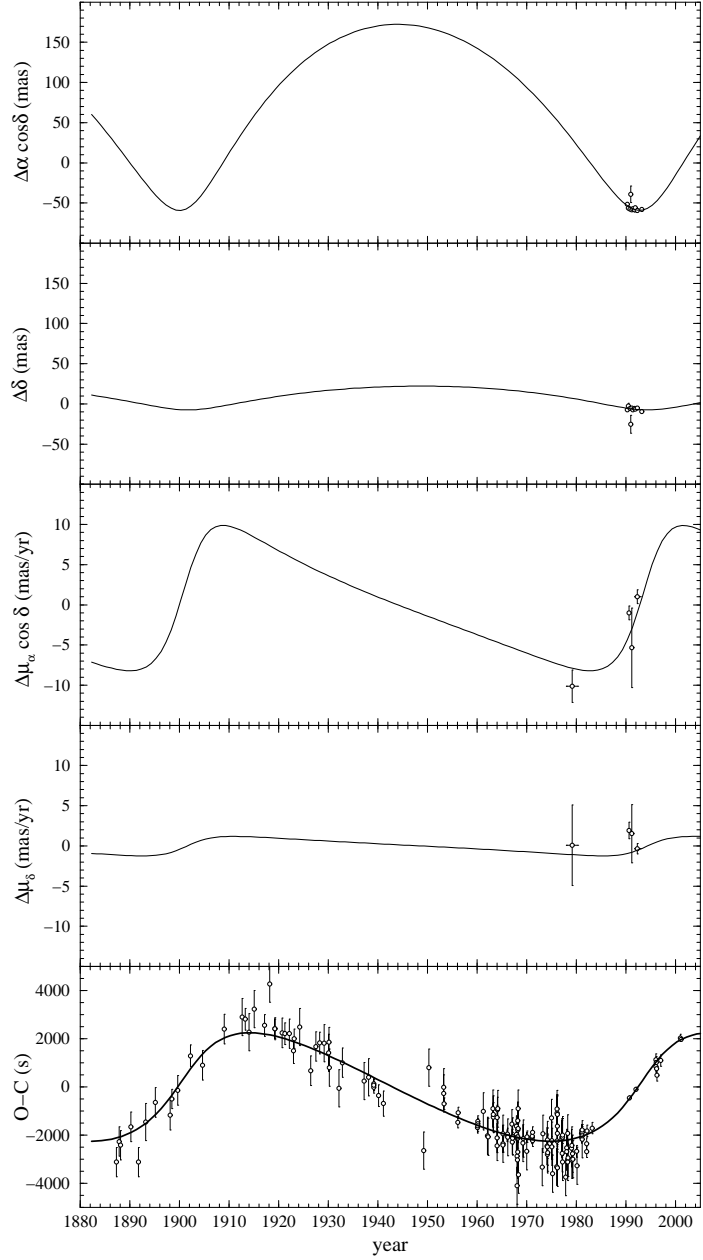}
\caption{Fits to the astrometric data (positions and proper
motions in the top 4 panels) and LTT curve (bottom panel) for R~CMa. Note that
the Hipparcos data are one-dimensional and thus cannot be represented directly.
Instead, we show the Hipparcos position for 10 normal points (epoch groups).
Although ground-based positions spanning over one century have been used to
constrain the least-square solution, these are not represented for clarity.
For the proper motions, the Guinan \& Ianna (1983) estimation and 3 Hipparcos
normal points are represented.
\label{figfit}}
\end{figure}

Due to the short timespan of the Hipparcos observations, a large uncertainty on
the reflex semi major-axis would exist if the Hipparcos astrometric data were
used alone. As a first step towards better constraining the solution we added
the epoch proper motion from Guinan \& Ianna (1983) as an external observation
(with the two subsequent equations for both components of the proper motion).
To do so, the appropriate equation for the first derivatives of the orbital
motion was used. As expected, the quality of the solution improved and yielded
a semi major axis of $a_{12}=140\pm16$ mas and a tertiary mass of
$M_3=0.42\pm0.05$ M$_{\odot}$. Yet, a closer inspection of the residuals
revealed that this solution was not fully compatible with the ground-based
epoch positions mentioned in \S \ref{astro}, as a clear trend appeared in
declination. For this reason we decided to include these positions
also in the fit, together with the Guinan \& Ianna proper motion and the
photometric (O--C) minimum times. In total, the least-squares fit had 262
equations for 14 parameters to determine. A robust fit approach (McArthur,
Jefferys, \& McCartney 1994) was used due to the large dispersion of the
ground-based astrometric measurements. The resulting goodness of the fit was
0.63 and graphical representations of the fits to the eclipse timing residuals
and Hipparcos intermediate astrometry are shown in Figure \ref{figfit}. The
less-accurate ground-based astrometric positions (with standard errors of about
200 mas on average) are not represented in the figure for the sake of clarity.

The resulting best-fitting parameters together with their standard deviations
are listed in Table \ref{tabprop}. The astrometric solution presented
supersedes that of the Hipparcos Catalogue because it is based upon a
sophisticated model that accounts for the orbital motion and considers
ground-based astrometry. Also, Table \ref{tabprop} includes the mass and
semi-major axis of R~CMa C that follow from the adoption of a total mass for
the eclipsing system. Finally, our fit also yields new accurate ephemeris for
the eclipsing pair:
\begin{equation} \label{ephem}
T(\mbox{Min I}) = {\rm HJED}2430436.5807 + 1.13594197 \; E
\end{equation}
where all times are in the TT scale and the zero epoch refers to the geometric
center of the R~CMa orbit. Note that the accuracy of the new period we
determine (see Table \ref{tabprop}) is better than 9 milliseconds. We have
considered in our analysis a linear ephemeris as that in Eq. (\ref{ephem}).
However, Algol systems have been observed to experience secular decreases of
the orbital periods possibly due to non-conservative mass transfer and angular
momentum loss (see, e.g., Qian 2000a). To assess the significance of this
effect on R CMa, we modified our fitting program by considering a quadratic
term. The coefficient of this quadratic term was found to be
$(-2.1\pm1.1)\cdot10^{-11}$ d, which translates into a period decrease rate of
$dP/dt = (-6.9\pm3.6)\cdot10^{-9}$ d yr$^{-1}$. This is a very slow rate
compared to other Algol systems (see, e.g., Qian 2001) yet commensurate with 
the low activity level of R CMa, which is near the end of its mass-transfer 
stage. Because of the poor significance (below 2$\sigma$) of the period 
decrease rate derived from the analysis, we decided to neglect the quadratic 
term and adopt a linear ephemeris. It should be pointed out, however, that 
the astrometric and orbital parameters resulting from the fit with quadratic 
ephemeris are well within one sigma of those listed in Table \ref{tabprop}.

\begin{figure}
\figurenum{2}
\epsscale{0.50}
\plotone{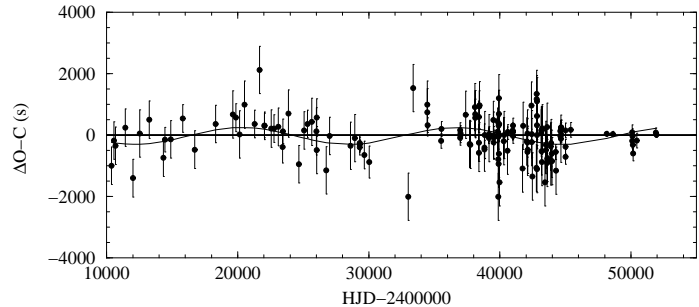}
\caption{Residuals of the fit to the LTT orbit of the third
body. The remaining oscillations have tentatively been modeled with an LTT
perturbation caused by a fourth body in a circular orbit. See text for fit
parameters.
\label{figres}}
\end{figure}

Interestingly, a closer inspection of Figure \ref{figfit} reveals small
excursions of the data from the LTT fit. To investigate these, we computed the
fit residuals which are shown in Figure \ref{figres}. The presence of
low-amplitude cyclic deviations seems quite obvious in this plot. If these
(O--C) timing oscillations were caused by the perturbation of a fourth body in
a circular orbit (R CMa D), its orbital period would be about 45 yr, with a LTT
semiamplitude of 275 s, a minimum mass of 0.06~M$_{\odot}$, and an orbital
semi-major axis of about 14 AU. The orbit of the third body is highly
eccentric so that it would be interior to that of R CMa C near its periastron
(${r_3}_{\rm P}=9.5$ AU). The intersections of the two orbits would result in
an apparently unstable configuration. Other possible explanations for the
low-amplitude oscillations include abrupt period changes of the binary itself
caused by variable angular momentum loss and magnetic coupling (see, e.g., Qian
2000b), a magnetic activity cycle of the secondary star (see, e.g., Applegate 
1992), or simply a spurious effect caused by the inhomogeneity of the data set. 
Unfortunately, the available astrometric data are not sensitive enough to prove 
or refute the existence of a fourth body and only new accurate photometric 
eclipse timing determinations or high-accuracy astrometry will provide the 
necessary evidence.

\section{Discussion}

The orbital, astrometric, and physical properties presented in Table
\ref{tabprop} are within 1$\sigma$ of the (less accurate) earlier estimates of
Radhakrishnan et al. (1984), who based their analysis on eclipse timings up to
1982. However, our study, in addition to extending the time baseline, has
been able to determine the inclination of the third body's orbit by making use
of the available high-precision astrometry (Hipparcos). Thus, R~CMa joins
Algol, the prototype of its class, in having the orbital properties of the
third body determined from a combined analysis of the astrometry and LTT.  The
long period ($\sim$93 yr) of R CMa C is the longest period detected and
confirmed so far for an eclipsing binary. This is chiefly because of the large
LTT present in R CMa (total amplitude of 86 min) and the existence of eclipse
timings available for this star back to 1887. 

It is interesting to note that the inclination of R CMa C is found to be of
$\approx$$92\pm5^{\circ}$ and thus compatible with an edge-on value of
90$^{\circ}$. In this situation, mutual eclipses of the tertiary component and
the close binary pair might occur. This tantalizing possibility is, however,
very unlikely since eclipses are only possible within a very narrow window
($\sim$1.2 arcmin) about an inclination of 90$^{\circ}$. If this were indeed 
the case, the transit of the tertiary component in front of the eclipsing pair
should have occurred during mid 2001. Also interesting to note is the near
coplanarity of the eclipsing system and its companion. Varricatt \& Ashok
(1999) found an inclination for the eclipsing pair of $i_{\rm EB}=79\fdg5$, 
which is equivalent to $i_{\rm EB}=100\fdg5$ because of the degeneracy. Thus, 
the third body's orbit appears to be within only 8--13$^{\circ}$ of the 
orbit of the eclipsing pair.

One question remains yet unaddressed, and this is the nature of the tertiary
companion of R~CMa. With a measured mass of 0.34~M$_{\odot}$, one is tempted to
classify R~CMa~C tentatively as a main sequence M3-4 star (Delfosse et al.
2000). However, another attractive possible scenario is a white dwarf (WD) as
tertiary component. There is no direct evidence for a WD companion to R~CMa,
but the mass of the third body is compatible with the low-end of the WD mass
distribution found by Silvestri et al. (2001). The presence of a hot WD
($T_{\rm eff}>10\,000$~K) is unlikely from IUE observations of R~CMa in the UV
region, where no hot source has been detected. Nonetheless, R~CMa is an old
disk population star so that a young WD is, in principle, not expected. If a WD
is present, its original stellar mass would have to be greater than the mass of
the initial primary (now secondary) of R~CMa. From binary evolution theory, the
best estimate of the initial mass of the original primary is about
$1.4$~M$_{\odot}$ (Sarma, Vivekananda Rao, \& Abhyankar 1996). This indicates a
pre-WD evolution time for the companion of around 2--3 Gyr. Cooling sequences
for WDs (Serenelli et al. 2001) yield an effective temperature of $\sim$5400~K
at an age of $\sim$3 Gyr, which is a reasonable estimate given the kinematic
characteristics of R~CMa. Should the tertiary component turn out to be a WD,
such an old and low-mass object might be exceedingly interesting since it could
belong to the controversial class of blue WDs that have been claimed to play an
important role in explaining the dark matter content of the galactic halo
(Hodgkin et al.  2000).

As the predicted temperatures both in the WD and M-star scenarios are fairly
similar, only the very different expected luminosities can help identify the
nature of the companion to R~CMa. Thus, the measure of the magnitudes of the
tertiary component through direct imaging would be a definitive proof. If we
consider the M-star scenario, the absolute magnitude of the tertiary component
would be $M_{\rm V}\approx11$~mag, which translates to $m_{\rm V}\approx14$~mag
when using the parallax obtained in \S\ref{analys}. This is about 8 mag fainter
than R~CMa itself. To give an example in the IR, the situation is significantly
improved in the K-band, where the magnitude difference is reduced to $\Delta
K\approx4$ mag. The tertiary component would be even fainter in the WD
scenario. Indeed, the absolute magnitude can be estimated as $M_{\rm
V}\approx14$~mag, which implies an apparent magnitude of $m_{\rm
V}\approx17$~mag. The difference with R CMa is therefore $\Delta
V\approx11$~mag. In the IR the situation does not improve significantly, with a
large magnitude difference of $\Delta K\approx10$~mag.

\begin{figure}
\figurenum{3}
\epsscale{0.50}
\plotone{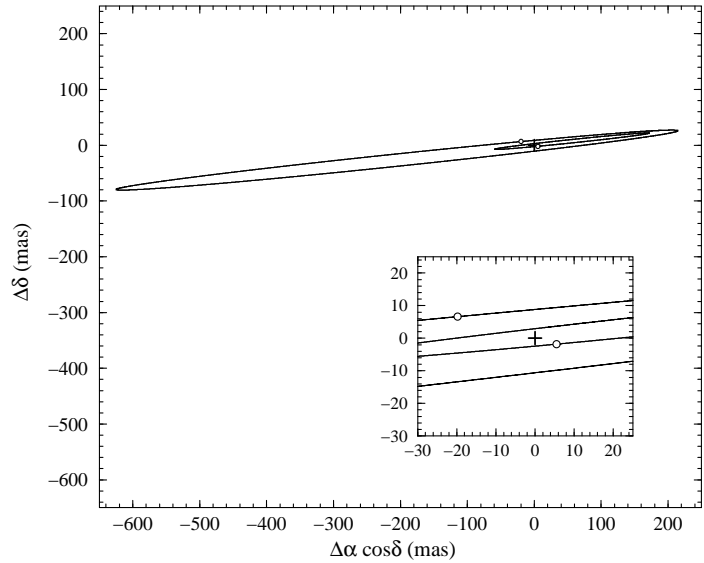}
\caption{Scale projection on the plane of the sky of the
orbits of the eclipsing pair of R~CMa (small ellipse) and the tertiary
component (large ellipse). The barycenter of the triple system is marked with a
plus sign and the positions of the stars in 2002 are represented as small
hollow circles. The inset shows a blowup of the region surrounding
the barycenter. The orbital properties and the sky projection were derived
from the simultaneous analysis of eclipse timing residuals and Hipparcos
astrometry (see \S\ref{analys}).
\label{figsky}}
\end{figure}

These magnitudes and dynamic brightness ranges are challenging but yet
attainable with state-of-the-art coronographs or Speckle spectrographs. Further
complications arise from the current spatial location of the tertiary component
near the conjunction of its orbit with the eclipsing pair.  Figure \ref{figsky}
depicts the predicted orbits of both the eclipsing pair and the tertiary
component on the plane of the sky. As can be seen, the separation between the
eclipsing system and R~CMa~C is only 27 mas as of 2002, which makes direct
imaging very difficult. On an optimistic note, the situation will slowly
improve in the future until a maximum separation of $\approx$$0\farcs8$ is
reached around year 2037.

Claims of third body detections through the analysis of (O--C) residuals have
sometimes been challenged. Spurious period changes caused by magnetic activity
cycles, variable angular momentum loss, magnetic coupling, or other effects
have been argued to explain modulations in the (O--C) residuals found in a
number of eclipsing binary stars. Interestingly, R~CMa would be a prime
candidate for such spurious period changes because of its interactive nature.
However, with over one period cycle in the LTT curve currently covered and,
more importantly, with direct evidence from Hipparcos astrometry, the case for
a tertiary companion to R~CMa is now iron-clad. What only remains to be
clarified at this point is whether this third star is an M dwarf or a WD. Also,
the nature of the lower amplitude $\sim$45 yr variation needs to be further
explored with continued observations.

\section{Conclusions}

This paper presents a combined analysis of short-term accurate astrometry and
long-term timing residuals applied to the eclipsing binary R~CMa. The study
yields the complete orbital and physical properties of the tertiary component. 
A determination of the mass of the third body is possible because the masses of
the eclipsing binary components themselves are well-known from light and radial
velocity curve analyses. 

The example discussed here illustrated the capabilities of a method that will
reach its full potential with the upcoming high-accuracy astrometric missions.
The improvements in precision of the future astrometric measurements are due to
an increase up to a thousand-fold relative to Hipparcos and the quality of the
photometry (and thus the eclipse timings) will also improve. More
quantitatively, timings with accuracies of $\sim$10 s are now possible for
select eclipsing binaries with sharp eclipses. The detection of large planets
($\sim$10~M$_{\rm J}$) in long-period orbits ($\sim$10--20~yr) around eclipsing
binaries will be therefore a relatively easy task. The short-term astrometry
will confirm the detections and yield the complete orbital solution (most
significantly the inclination) and thus the actual mass of the orbiting body.

One of the unexpected outcomes of the Hipparcos mission has been that a
primarily astrometric satellite can also provide valuable new results from its
photometric measurements alone (numerous new variables, HD 209458 planetary
transits, etc). The data analysis of the next generations of astrometric
satellites will surely benefit from a simultaneous analysis of the astrometric
and photometric data.  Astrometric missions such as GAIA will likely detect one
million new eclipsing binaries (a smaller number is expected for FAME). About
one per cent of the eclipsing binaries observed by Hipparcos has a 0.0001 day
precision in the reference epoch, which is enough to detect the LTT effect that
would arise from a 10 Jupiter mass third body with a 11 year period. If we
assume the same ratio for GAIA, hundreds to thousands of third bodies would be
detected. Although GAIA astrometry alone will be able to give the orbit for the
closest stars, the orbit for more distant stars will depend on the availability
of ground-based light curves to define the reference epoch.

This method of combining LTT analysis and astrometry complements very
well with the ongoing spectroscopic searches. The LTT analysis favors the
detection of long-period third bodies around eclipsing binaries because the
amplitude of the time delay due to the LTT effect is proportional to
$P_{12}^{2/3}$ while the spectroscopic semi-amplitude is proportional to
$P_{12}^{-1/3}$.  When the samples of spectroscopic and LTT systems are
sufficiently large, we will have a complete picture of the distribution of
bodies in a stellar system and a realistic test of planet formation theories
will be possible.

Finally, the LTT analysis method does not have to be necessarily applied to
eclipsing binaries. In essence, the method is based upon having a ``beacon in
orbit'', which, in the case of eclipsing binaries, are the mid-eclipse times.
However, any strictly periodic event that can be predicted with good accuracy
could be potentially useful to detect stellar or sub-stellar companions. This
includes, for example, pulsating stars. More interestingly, transiting planets
are also prime candidates for LTT studies. In this case, not only further
orbiting planets could be discovered, but also good chances for detecting moons
around the transiting planet exist.

\acknowledgments

Dr. S. E. Urban (US Naval Observatory), who provided the older epoch
astrometric data of R~CMa is warmly thanked. The APT observations were acquired
and reduced using programs developed by Dr. G. P. McCook, who is gratefully
acknowledged. Also, we thank astronomy students J. J. Bochanski and D. Stack
for help with data preparation for this work. The anonymous referee is thanked
for a number of important comments and suggestions that led to the improvement
of the paper. I. R. thanks the Catalan Regional Government (CIRIT) for
financial support through a Fulbright fellowship. This research was supported
by NSF/RUI grants AST 93-15365, AST 95-28506, and AST 00-71260.

\begin{deluxetable}{lllrrl}
\tablewidth{0pt}
\tablefontsize{\footnotesize}
\tablecaption{Ground-based astrometric data for R CMa.
\label{tabastro}}
\tablehead{\colhead{Julian year}&
\colhead{$\alpha$ (deg)}& 
\colhead{$\delta$ (deg)}&
\colhead{$\sigma_\alpha\cos\delta$ (mas)}&
\colhead{$\sigma_\delta$ (mas)}&
\colhead{Source}}
\startdata
1894.7     &109.86244875  &$-$16.391100556&       861                       &       690           & WASH AG 1900      \\
1905.15    &109.8630725   &$-$16.391588056&       253                       &       269           & A.C.              \\
1914.11    &109.863289583 &$-$16.391985556&       132                       &       116           & CAPE 2ND FUND 1900\\
1914.97    &109.863497083 &$-$16.391931111&       304                       &       332           & A.C.              \\
1916.2     &109.863377917 &$-$16.392131667&       294                       &       282           & ALBANY 10         \\
1918.5     &109.863659583 &$-$16.392145833&       321                       &       270           & WASH 20           \\
1923.05    &109.863762083 &$-$16.392368333&       156                       &       153           & CAPE 1-25         \\
1933.52    &109.864227083 &$-$16.392731944&       316                       &       264           & YALE 12/1 -14/-18 \\
1934.44    &109.86428875  &$-$16.392647222&       237                       &       210           & CAPE 3-25         \\
1939.58    &109.86449     &$-$16.392880556&       130                       &       132           & CAPE 1-50         \\
1942.02    &109.864635833 &$-$16.393026944&       151                       &       200           & WASH 40 9-IN      \\
1969.33    &109.865958333 &$-$16.394058889&        45                       &        45           & CPC2              \\
1983.05    &109.866597083 &$-$16.394559722&       168                       &       164           & WASH TAC          \\
1984.37    &109.866713333 &$-$16.394581389&        86                       &       181           & PERTH 83          \\
1986       &109.8667525   &$-$16.394728333&        88                       &       119           & CAMC Series       \\
1986.07    &109.866875417 &$-$16.394721944&       110                       &       119           & FOKAT             \\
1992.49    &109.8670625   &$-$16.394969444&        39                       &        46           & WASH 2-J00        \\
\enddata

References for Table \ref{tabastro}: 
WASH -- Urban (2001; priv. comm.);
A.C. -- Urban et al. (2000);
CAPE -- Urban (2001; priv. comm.);
ALBANY -- Urban (2001; priv. comm.);
YALE -- Urban (2001; priv. comm.);
CPC2 -- Zacharias et al. (1992); 
PERTH -- Urban (2001; priv. comm.);
CAMC -- Fabricius (1993); 
FOKAT -- Bystrov et al. (1991). 
\end{deluxetable}

\begin{deluxetable}{crrrcrrrcrrr}
\tablewidth{0pt}
\tablefontsize{\footnotesize}
\tablecaption{Primary eclipse timings for R CMa. 
\label{tabtim}}
\tablehead{
\colhead{HJD\tablenotemark{a}} &
\colhead{(O--C) (s)} & 
\colhead{$\sigma$\tablenotemark{b} (s)} &
\colhead{Ref.} &
\colhead{HJD\tablenotemark{a}} &
\colhead{(O--C) (s)} & 
\colhead{$\sigma$\tablenotemark{b} (s)} &
\colhead{Ref.} &
\colhead{HJD\tablenotemark{a}} &
\colhead{(O--C) (s)} & 
\colhead{$\sigma$\tablenotemark{b} (s)} &
\colhead{Ref.}} 
\startdata
2410368.9939&$-$3114&610&1&2436982.9957&$-$1608&240&1&2442402.5785                 &$-$1287&770&1\\
2410562.1139&$-$2262&610&1&2437378.3104&$-$1012&770&4&2442426.4195                 &$-$2477&770&1\\
2410664.3469&$-$2415&610&1&2437696.3624&$-$2027&770&4&2442426.4225                 &$-$2218&770&1\\
2411425.4369&$-$1648&610&1&2437746.3434&$-$2065&770&4&2442467.3005                 &$-$3593&770&1\\
2411993.3909&$-$3116&610&1&2438089.4114& $-$896&770&4&2442785.3675                 &$-$3311&770&1\\
2412527.3029&$-$1451&770&1&2438105.3104&$-$1258&610&4&2442802.4325                 &$-$1076&770&1\\
2413242.9557& $-$642&610&1&2438114.3994&$-$1132&770&4&2442802.4345                 & $-$903&770&1\\
2414333.4539&$-$1166&610&1&2438384.7384&$-$2443&610&4&2442820.5941                 &$-$2240&240&1\\
2414447.0559& $-$491&400&1&2438399.5192&$-$1272&520&4&2442826.2775                 &$-$1921&770&1\\
2414878.7180& $-$140&610&1&2438400.6454&$-$2114&610&4&2442826.2865                 &$-$1143&770&1\\
2415810.2070&   1296&450&1&2438406.3387& $-$940&520&4&2442835.3485                 &$-$3350&770&1\\
2416718.9560&    904&610&2&2438440.4174& $-$902&770&4&2442835.3685                 &$-$1622&770&1\\
2418309.2921&   2400&610&1&2438817.5334&$-$2347&770&4&2443161.3775                 &$-$2169&770&1\\
2419615.6312&   2901&770&1&2438818.6687&$-$2402&290&4&2443162.5145                 &$-$2077&770&1\\
2419849.6342&   2811&450&1&2438832.3054&$-$1936&770&4&2443186.3616                 &$-$2750&770&1\\
2420138.1572&   2271&770&1&2439140.1446&$-$2028&240&1&2443202.2725                 &$-$2075&770&1\\
2420513.0292&   3236&770&1&2439164.0002&$-$1957&240&1&2443203.3966                 &$-$3106&770&1\\
2421278.6462&   2557&450&1&2439169.6784&$-$2087&770&1&2443219.3125                 &$-$1999&770&1\\
2421648.9832&   4279&770&1&2439492.2922&$-$1544&610&5&2443430.5776                 &$-$3744&770&1\\
2422029.5022&   2417&450&1&2439518.4104&$-$2275&520&4&2443512.3796                 &$-$2519&770&1\\
2422030.6382&   2422&400&1&2439528.6364&$-$2057&770&4&2443513.5136                 &$-$2687&770&1\\
2422558.8492&   2249&610&1&2439533.1794&$-$2124&240&1&2443587.3586                 &$-$1929&770&1\\
2422765.5903&   2211&450&1&2439802.4034&$-$1626&240&1&2443595.2966                 &$-$3103&770&1\\
2423098.4213&   2212&610&1&2439822.8464&$-$1967&770&1&2443612.3376                 &$-$2942&770&1\\
2423406.2533&   1498&520&1&2439863.7384&$-$2132&770&1&2443880.4246                 &$-$2535&770&1\\
2423442.6093&   2004&400&1&2439870.5314&$-$4090&770&1&2443888.3706                 &$-$3018&770&1\\
2423866.3213&   2492&770&1&2439872.8174&$-$2870&770&1&2443905.4166                 &$-$2425&770&1\\
2424667.1393&    671&610&1&2439875.0984&$-$2082& 80&1&2443946.3066                 &$-$2763&770&1\\
2425052.2353&   1679&610&1&2439896.6704&$-$3024&770&1&2443971.2946                 &$-$2998&770&1\\
2425320.3193&   1825&450&1&2439904.6324&$-$2125&770&1&2444255.2839                 &$-$2668&240&1\\
2425650.8783&   1816&770&1&2439905.7724&$-$1774&770&1&2444281.4036                 &$-$3270&770&1\\
2425990.5203&   1415&770&1&2439912.5774&$-$2694&770&1&2444606.2986                 &$-$1921&240&1\\
2426014.3803&   1866&610&1&2439912.5847&$-$2064&240&1&2444607.4327                 &$-$2080&240&1\\
2426027.9993&    803&770&1&2439912.5924&$-$1398&770&1&2444647.1938                 &$-$1810&240&1\\
2426753.8563&  $-$53&770&1&2439929.6374& $-$891&770&1&2444648.3289                 &$-$1883&240&1\\
2426994.6883&   1009&610&1&2439935.3075&$-$1730&240&1&2444649.4616                 &$-$2163&770&1\\
2428596.3576&    242&770&1&2439954.5965&$-$3632&770&1&2444664.2304                 &$-$2028&240&1\\
2428922.3748&    403&770&1&2440288.5785&$-$2330&770&1&2444672.1848                 &$-$1786&240&1\\
2429301.7763&    133&240&1&2440313.5715&$-$2133&770&1&2444998.1898                 &$-$2679&240&1\\
2429308.5903&   $-$8&240&1&2440582.7835&$-$2672&770&1&2444999.3295                 &$-$2354&240&1\\
2429309.7273&     82&240&1&2440591.8781&$-$2061&240&6&2445015.2389                 &$-$1817&240&1\\
2429660.7283& $-$355&450&1&2440964.4665&$-$2109&240&1&2445391.2370                 &$-$1704&240&7\\
2430035.5853& $-$687&520&1&2440971.2825&$-$2079&240&1&2448137.9592\tablenotemark{c}& $-$449& 80&8\\
2432999.2353&$-$2634&770&1&2440979.2345&$-$2044&240&1&2448608.2433\tablenotemark{c}&  $-$93& 80&8\\
2433367.3203&    804&770&1&2440995.1395&$-$1887&240&1&2450088.3866                 &    850&240&9\\
2434453.2714&  $-$16&770&1&2440996.2715&$-$2228&240&1&2450096.3415                 &   1136&240&9\\
2434454.4043& $-$270&770&1&2441725.5335&$-$3327&770&1&2450107.6995                 &   1013&240&8\\
2434481.6620& $-$694&240&1&2441765.3075&$-$1941&770&1&2450145.1826                 &    756&240&9\\
2435515.3604&$-$1461&240&1&2442059.5105&$-$2456&770&1&2450154.2670                 &    485&240&9\\
2435534.6759&$-$1074&240&1&2442092.4525&$-$2484&770&1&2450439.3955                 &   1096&240&9\\
2436958.0042&$-$1675&240&1&2442099.2715&$-$2194&770&1&2451896.8199                 &   2036&150&8\\
2436959.1430&$-$1428&240&1&2442100.4005&$-$2794&770&1&2451945.6648                 &   1977& 80&8\\
2436977.3169&$-$1530&240&3&2442116.3045&$-$2724&770&1&\nodata&\nodata&\nodata&\nodata\\            
\enddata
\tablenotetext{a}{Not in the UTC but in the TT scale (see text).}
\tablenotetext{b}{The uncertainties have been computed using the procedure
outlined in \S\ref{secphot}.}
\tablenotetext{c}{From Hipparcos photometry.}
References for Table \ref{tabtim}: 
(1) Radhakrishnan et al. (1984);         
(2) Wood (1946);                         
(3) Knipe (1963);                        
(4) Kitamura (1969);                     
(5) Robinson (1967);                     
(6) Guinan (unpub.);                     
(7) Edalati, Khalesse, \& Riazi (1989);  
(8) This work;                           
(9) Varricatt \& Ashok (1999).           
\end{deluxetable}

\begin{deluxetable}{lc}
\tablewidth{0pt}
\tablefontsize{\footnotesize}
\tablecaption{Astrometric and light-travel time (LTT) solutions for the triple
system R~CMa. \label{tabprop}}
\tablehead{\colhead{Parameter}&\colhead{Value \& Standard 
error}}
\startdata
$\pi$ (mas)                                 &   22.70$\pm$0.89  \\
$\mu_{\alpha} \cos \delta$ (mas yr$^{-1}$)  &   168.1$\pm$0.7   \\
$\mu_{\delta}$ (mas yr$^{-1}$)              &$-$137.1$\pm$1.2   \\
$a_{12}$ (mas)                              &   117.2$\pm$5.3   \\
$\Omega_{12}$ (\arcdeg)                     &   262.9$\pm$20.7  \\
$P_{12}$ (yr)                               &    92.8$\pm$1.3   \\
$e_{12}$                                    &    0.49$\pm$0.05  \\
$i_{12}$ (\arcdeg)                          &    91.7$\pm$4.7   \\
$\omega_{12}$ (\arcdeg)                     &    10.5$\pm$4.3   \\
$T_{12}^{\rm peri}$ (HJED)                  & 2449343$\pm$258   \\
LTT semiamp. (s)                            &    2574$\pm$57    \\
$M_{12}$ (M$_{\odot}$)                      &  1.24$\pm$0.05\tablenotemark{a}\\
$M_3$ (M$_{\odot}$)                         &  0.34$\pm$0.02    \\
$a_3$ (AU)                                  &  18.7$\pm$1.7     \\
$P_{\rm EB}$ (d)                            &  1.13594197$\pm$0.00000010 \\
${T_{\circ}}_{\rm EB}$ (HJED)               &  2430436.5807$\pm$0.0006 \\
\enddata
\tablenotetext{a}{Adopted from Tomkin 1985.}
\end{deluxetable}

\end{document}